\documentclass[11pt]{article}

\usepackage{amsmath}
\usepackage{graphicx}
\usepackage{indentfirst}
\usepackage{amssymb}
\usepackage{cite}
\usepackage{color}
\usepackage{subfigure}

\begin{document}

\title{\bf{Strong Cosmic Censorship under Quasinormal Modes of Non-Minimally Coupled Massive Scalar Field}}

\date{}
\maketitle

\begin{center}
\author{Bogeun Gwak}$^a$\footnote{rasenis@dongguk.edu}\\

\vskip 0.25in
$^{a}$\it{Division of Physics and Semiconductor Science, Dongguk University, Seoul 04620,\\Republic of Korea}\\
\end{center}
\vskip 0.6in

{\abstract
{We investigate the strong cosmic censorship conjecture in lukewarm Reissner-Nordstr\"{o}m-de Sitter black holes (and Mart\'{i}nez-Troncoso-Zanelli black holes) using the quasinormal resonance of non-minimally coupled massive scalar field. The strong cosmic censorship conjecture is closely related to the stability of the Cauchy horizon governed by the decay rate of the dominant quasinormal mode. Here, dominant modes are obtained in the limits of small and large mass black holes. Then, we connect the modes by using the WKB approximation. In our analysis, the strong cosmic censorship conjecture is valid except in the range of the small-mass limit, in which the dominant mode can be assumed to be that of the de Sitter spacetime. Particularly, the coupling constant and mass of the scalar field determine the decay rate in the small mass range. Therefore, the validity of the strong cosmic censorship conjecture depends on the characteristics of the scalar field.
}}

\thispagestyle{empty}
\newpage
\setcounter{page}{1}

\section{Introduction}\label{sec1}

The inside of a black hole is covered by its event horizon, from which no light can escape. Hence, it is not possible to detect black holes by its own radiation, classically. However, quantum theory suggests that a black hole can emit a portion of the energy from the horizon. This radiation is called Hawking radiation\cite{Hawking:1974sw,Hawking:1976de}. In consideration of Hawking radiation, a black hole can be treated as a thermodynamic system with the Hawking temperature, which is proportional to the surface gravity at the horizon. Moreover, the area of the  black hole¡¯s horizon is irreducible in an irreversible process\cite{Christodoulou:1970wf,Christodoulou:1972kt,Smarr:1972kt}. Therefore, Bekenstein-Hawking entropy of a black hole is defined as being proportional to the area of its event horizon \cite{Bekenstein:1973ur,Bekenstein:1974ax}. Because properties of black holes are different from any other astronomical object in the universe, the existence of the black hole evokes curiosity. However, the recent detection of gravitational wave signals, which originated from collisions between black holes, by the Laser Interferometer Gravitational-Wave Observatory (LIGO) has proven that black holes are in fact stable celestial bodies spread across the universe.

The center of a black hole is the location of a curvature singularity. Physically, a visible singularity causes the breakdown of causality and loss of predictability in the theory of gravity. Hence, to avoid this unpredictability, the singularity should be invisible to the observer. This is called the cosmic censorship conjecture\cite{Penrose:1964wq,Penrose:1969pc,Hawking:1969sw}. According to a given observer, the cosmic censorship conjecture is divided into two types: weak conjecture and strong conjecture. On the one hand, the weak cosmic censorship (WCC) conjecture states that the singularity should be covered by an outer horizon for an asymptotic observer. Thus, the outer horizon needs to be stable under perturbation to satisfy the WCC conjecture. The first test on the WCC conjecture was performed on the Kerr black hole\cite{Wald:1974ge}. Here, adding a particle into the Kerr black hole cannot overspin it beyond the extremality. Since then, the WCC conjecture has been tested in various black holes. Moreover, the validity of this conjecture depends on the state of the black hole and the method of perturbation. For example, the horizon of the near-extremal Kerr black hole becomes unstable upon adding a particle\cite{Jacobson:2009kt}, but it can be still be stable when considering self-force effects\cite{Barausse:2010ka,Barausse:2011vx,Colleoni:2015ena,Colleoni:2015afa,Sorce:2017dst}. This test can be extended to the Reissner-Nordstr\"{o}m black hole where a situation similar to that of the Kerr black hole arises, as discussed in \cite{Hubeny:1998ga,Isoyama:2011ea}. Further, as there is no general proof of the validity of the WCC conjecture, the test is now extended to various black holes by adding a particle\cite{Crisostomo:2003xz,Gwak:2011rp,Gao:2012ca,Hod:2013vj,Rocha:2014jma,Gwak:2015fsa,Cardoso:2015xtj,Horowitz:2016ezu,Revelar:2017sem,Duztas:2017lxk,Yu:2018eqq,Shaymatov:2018fmp,Gim:2018axz}. Particularly, when the thermodynamic pressure and volume terms are considered for the electrically charged anti-de Sitter black hole, the WCC conjecture is proven to be valid under particle absorption\cite{Gwak:2017kkt}. For the test of the WCC conjecture, adding a particle can be generalized to the scattering of the test field\cite{Hod:2008zza,Semiz:2005gs,Toth:2011ab,Duztas:2013wua,Semiz:2015pna,Natario:2016bay,Duztas:2018adf,Chen:2018yah}. Under the scattering of a scalar field, the weak cosmic censorship conjecture is shown to be valid for the Kerr-(anti-)de Sitter black holes\cite{Gwak:2018akg}.

The strong cosmic censorship (SCC) conjecture proposes that the singularity is invisible to any observer, and hence must be a spacelike singularity. It is important to note that a timelike singularity appears in well-known solutions such as Reissner-Nordstr\"{o}m and Kerr-Newman black holes, which often leads to the notion that these black holes are counterexamples to the SCC conjecture. However, this is not true because inside the outer horizon, the timelike singularity is enclosed by a Cauchy (inner) horizon at which an inward wave undergoes an infinite blueshift. Hence, when a wave enters the black hole, the infinitely blueshifted wave makes the Cauchy horizon unstable. As a result, the singularity becomes spacelike, making the SCC conjecture valid even in this case\cite{Matzner:1979zz,Poisson:1990eh,Ori:1991zz,Brady:1995ni,Konoplya:2011qq,Konoplya:2018qov}. However, the issues with the SCC conjecture becomes more complicated in the Reissner-Nordstr\"{o}m-de Sitter (RNdS) black hole. For instance, in the de Sitter (dS) spacetime, the existence of a cosmological horizon causes the redshift of an influx into the Cauchy horizon, so the redshift competes with the blueshift from the Cauchy horizon. Then, the redshift originating from the cosmological horizon becomes dominant, stabilizing the Cauchy horizon \cite{Mellor:1989ac}. However, there can be an additional influx to the Cauchy horizon. Furthermore, this influx is predominantly blueshifted at the Cauchy horizon, which can destabilize the Cauchy horizon \cite{Brady:1998au}. Recently, quasinormal modes have been physically categorized into three families based on their behaviors in an RNdS black hole, and these behaviors play an important role in the validity of the SCC conjecture\cite{Burikham:2017gdm,Cardoso:2017soq}. Particularly, the stability of the Cauchy horizon depends significantly on the competition between the surface gravity on the Cauchy horizon and the decay rate of the quasinormal mode on the outer horizon. By the analysis of the quasinormal modes in RNdS black holes, ranges have been found over which the SCC conjecture is invalid\cite{Cardoso:2017soq}. Nevertheless, the SCC conjecture in the RNdS black hole is still actively studied in \cite{Hod:2018dpx,Cardoso:2018nvb,Dias:2018etb,Mo:2018nnu,Dias:2018ufh,Luna:2018jfk,Ge:2018vjq,Destounis:2018qnb,Rahman:2018oso}. The historical review can be found in \cite{Chambers:1997ef} (and references therein).

Here, we consider a dS black hole whose metric is a solution (of the same geometry) to two theories of gravity: Einstein's gravity coupled with the Maxwell field, and gravity theory coupled with a conformal scalar field including a quartic self-interaction potential. The solution in these two theories is known by different names. The former is called the lukewarm RNdS black hole\cite{Romans:1991nq} and the latter Mart\'{i}nez-Troncoso-Zanelli (MTZ) black hole\cite{Martinez:2002ru}. Further, dS black holes encounter an issue with the temperature. Since dS black holes have two horizons surrounding the timelike spacetime, two temperatures for the two horizons can be obtained. It should be noted that these two temperatures are not coincident, so the system is not balanced between the input and output radiations through the horizons. Hence, the systems are thermodynamically unstable. The lukewarm RNdS black hole resolves unbalanced radiations by setting the two temperatures at a coinciding value\cite{Romans:1991nq,Cai:1997ih,Mann:1995vb}. In the gravity theory coupled with a conformal scalar field including a quartic self-interaction potential, the geometry becomes that of the MTZ black hole, which is a four-dimensional dS black hole with a non-singular scalar hair outside the outer horizon. Further, the MTZ black hole can satisfy the strong energy condition\cite{Martinez:2002ru}. However, during a perturbation, an instability can be observed in the MTZ black hole\cite{Harper:2003wt}, which is consistent with the no-hair theorem. Thermodynamically, according to the effect of the scalar field, derived from the Euclidean
action, the entropy of the MTZ black hole is given by a modified form\cite{Barlow:2005yd}.

In this work, we investigate the SCC conjecture in the lukewarm RNdS (or MTZ) black hole under the quasinormal modes of non-minimally coupled massive scalar field. The decay rate of the scalar field is closely related to the investigation of the SCC conjecture. In our analysis, as the decay rate depends on non-minimal coupling and scalar field mass, we elucidate these effects in the SCC conjecture, which has not been done yet under the non-minimally coupled massive scalar field. Further, in the case of the lukewarm RNdS black hole, we will investigate the SCC conjecture for a thermally stable dS black hole and test its consistency with previous studies on non-lukewarm RNdS black holes. In the case of the MTZ black hole, the SCC conjecture for hairy black holes has not been studied much. Although the MTZ black hole is unstable, we propose it is a useful solution to extending studies on its SCC conjecture to black holes having scalar hair. It should be noted that because our analysis is based on the quasinormal resonances that are considered linear effects of the scalar field, its results depend only on the equations of motion for the scalar field rather than on the action for gravity theories. Therefore, our conclusion on the SCC conjecture are the same for both the black holes. Here, for convenience, we will call the geometry as the MTZ black hole.

The paper is organized as follows: Section\,\ref{sec2} introduces the geometry of the MTZ black hole. Section\,\ref{sec3} solves the non-minimally coupled massive scalar field equation at the outer horizon in the MTZ black hole. Section\,\ref{sec4} investigates the SCC conjecture in two limits of the scalar field's mass. Then, it approximates the quasinormal modes in the intermediate range of the mass by the WKB method. Section\,\ref{sec6} summarizes the results.

\section{Geometry of dS Black Holes}\label{sec2}

The spacetime geometry, as we consider, is a dS black hole. The metric is given as
\begin{align}\label{eq:metric01}
ds^2 = - \frac{\Delta}{r^2}dt^2+ \frac{r^2}{\Delta}dr^2+r^2 d\theta^2 +r^2 \sin^2\theta d\phi^2,\quad \Delta=-\frac{\Lambda r^4}{3}+(r-M)^2,
\end{align}
which is defined as a black hole having mass $M$ and cosmological constant $\Lambda$. The curvature singularity is located at the center of the spacetime. In the limit of the asymptotic region, the metric becomes the dS spacetime containing the cosmological horizon. The mass of the black hole is in the range of $0<M<\frac{1}{4}\sqrt{\frac{3}{\Lambda}}$. There exists four solutions to $g^{rr}=\Delta(r)=0$ in the spacetime
\begin{align}
r_\text{i}&= \frac{1}{2}\sqrt{\frac{3}{\Lambda}}\left(-1+\sqrt{1+4M\sqrt{\frac{\Lambda}{3}}}\right),\quad r_\text{o}= \frac{1}{2}\sqrt{\frac{3}{\Lambda}}\left(1-\sqrt{1-4M\sqrt{\frac{\Lambda}{3}}}\right),\\
r_\text{c}&= \frac{1}{2}\sqrt{\frac{3}{\Lambda}} \left(1+\sqrt{1-4M\sqrt{\frac{\Lambda}{3}}}\right),\quad r_{\rm n}=-\frac{1}{2}\sqrt{\frac{3}{\Lambda}}\left(1+\sqrt{1+4M\sqrt{\frac{\Lambda}{3}}}\right),\nonumber
\end{align}
where $r_\text{i}$, $r_\text{o}$, and $r_\text{c}$ correspond to Cauchy (inner), outer, and cosmological horizons, and $r_\text{n}$ has no physical correspondence. Note that $G=1$ in this case. In our analysis of the SCC conjecture, the surface gravities on the inner and outer horizons, $\kappa_\text{i}$ and $\kappa_\text{o}$, play important roles in competing with the amplification and decay rates of the scalar field. Then,
\begin{align}
\kappa_\text{i}=\frac{1}{2}\left|\frac{d}{dr}\left(\frac{\Delta}{r^2}\right)\right|_{r=r_{\rm i}}
= \sqrt{\frac{\Lambda}{3}}\sqrt{1+4M\sqrt{\frac{\Lambda}{3}}},\quad \kappa_\text{o} =\frac{1}{2}\left|\frac{d}{dr}\left(\frac{\Delta}{r^2}\right)\right|_{r=r_{\rm o}}
= \sqrt{\frac{\Lambda}{3}}\sqrt{1-4M\sqrt{\frac{\Lambda}{3}}}.
\end{align}
Interestingly, the metric of Eq.\,(\ref{eq:metric01}) appears in the same form in both Einstein-Maxwell action and gravity action coupled with conformal scalar field that includes a quartic self-interaction potential. Hence, the geometric properties coincide as reviewed above. However, the physics of the two, such as coupling fields, is different from each other. Therefore, we introduce them as follows.

\subsection{Lukewarm Reissner-Nordstr\"{o}m-de Sitter Black Hole}

The Lukewarm RNdS black hole is the solution to the Einstein-Maxwell action with the cosmological constant
\begin{align}\label{eq:EMCaction1}
S=\frac{1}{16\pi}\int d^4 x \sqrt{-g} \left(R-F_{\mu\nu}F^{\mu\nu}-2\Lambda\right).
\end{align}
$F_{\mu\nu}$ and $A_\mu$ are the Maxwell field strength and electric potential of a charge $Q$ related to
\begin{align}
F_{\mu\nu}=\partial_\mu A_\nu - \partial_\nu A_{\mu}, \quad  A=-\frac{Q}{r}dt.
\end{align}
The field equations of Eq.\,(\ref{eq:EMCaction1}) contain a spherical symmetric solution to the RNdS black hole whose the metric is obtained as
\begin{align}\label{eq:metric02rn}
ds^2 = - \frac{\Delta}{r^2}dt^2+ \frac{r^2}{\Delta}dr^2+r^2 d\theta^2 +r^2 \sin^2\theta d\phi^2,\quad \Delta=-\frac{\Lambda r^4}{3}+r^2-2Mr+Q^2.
\end{align}
Here, the Hawking temperatures on the outer and cosmological horizons in Eq.\,(\ref{eq:metric02rn}) are
\begin{align}
T_\text{o}=\frac{1}{2\pi}\left(-\frac{Q^2}{r_\text{o}^3}+\frac{M}{r_\text{o}^2}-\frac{r_\text{o}\Lambda}{3}\right),\quad T_\text{c} =\frac{1}{2\pi}\left(-\frac{Q^2}{r_\text{c}^3}+\frac{M}{r_\text{c}^2}-\frac{r_\text{c}\Lambda}{3}\right),
\end{align}
in which the difference between the two temperatures implies that radiations are not in equilibrium. Hence, the thermodynamic system is unstable. The RNdS black hole can be in thermal equilibrium when the two temperatures become equal to one another. Equal temperatures are achieved at the same mass and electric charge, $M=Q$. Then,
\begin{align}
T_\text{o}=T_\text{c}=\frac{1}{2\pi}\sqrt{\frac{\Lambda}{3}}\sqrt{1-4M\sqrt{\frac{\Lambda}{3}}}.
\end{align}
This is called the lukewarm RNdS black hole\cite{Romans:1991nq}, and its metric is exactly as that in Eq.\,(\ref{eq:metric01}). Note that the lukewarm RNdS black hole has only an electric charge $Q=M$ so that it can still be coupled with an external electric charge.

\subsection{Mart\'{i}nez-Troncoso-Zanelli Black Hole}

The MTZ black hole appears in the four-dimensional theory of gravity coupled with conformal scalar field including a quartic self-interaction potential\cite{Martinez:2002ru}. The action is
\begin{align}
S=\int d^4 x \sqrt{-g}\left(\frac{\mathcal{R}-2\Lambda}{16\pi G}-\frac{1}{2}\partial_\mu \Phi \partial^\mu \Phi-\frac{1}{12}\mathcal{R}\Phi^2 -\alpha \Phi^4\right),
\end{align}
where $\alpha$ is a dimensionless constant. The MTZ black hole is a solution to the field equations
\begin{align}
G_{\mu\nu}+\Lambda g_{\mu\nu} = 8 \pi G T_{\mu\nu},\quad \Box \Phi -\frac{1}{6}\mathcal{R}\Phi -4\alpha \Phi^3=0,
\end{align}
where the energy-momentum tensor is given by
\begin{align}
T_{\mu\nu}=\partial_\mu \Phi \partial_\nu \Phi -\frac{1}{2}g_{\mu\nu}\partial_\sigma \Phi \partial^\sigma \Phi + \frac{1}{6}(g_{\mu\nu}\Box-\nabla_\mu \nabla_\nu+G_{\mu\nu})\Phi^2-\alpha g_{\mu\nu}\Phi^4.
\end{align}
When the parameter pair is chosen as
\begin{align}
\mathcal{R}=4\Lambda,\quad \alpha=-\frac{2}{9}\pi \Lambda,
\end{align}
the MTZ black hole is the only solution expressed as Eq.\,(\ref{eq:metric01}) in a positive cosmological constant\cite{Martinez:2002ru}. Given that the action is coupled with a scalar field, the MTZ black hole includes a scalar hair obtained as
\begin{align}
\Phi(r)=\sqrt{\frac{3}{4\pi}} \frac{M}{r-M},
\end{align}
which is non-singular outside the outer horizon. Compared with the RNdS black hole, the MTZ black hole is neutral. Hence, the system is not coupled with an electric charge.

\section{Non-Minimally Coupled Massive Scalar Field}\label{sec3}

\begin{figure}[h]
\centering
\subfigure[{Effective potentials in $\mu = 1$ and $\xi = 1$.}] {\includegraphics[scale=0.6,keepaspectratio]{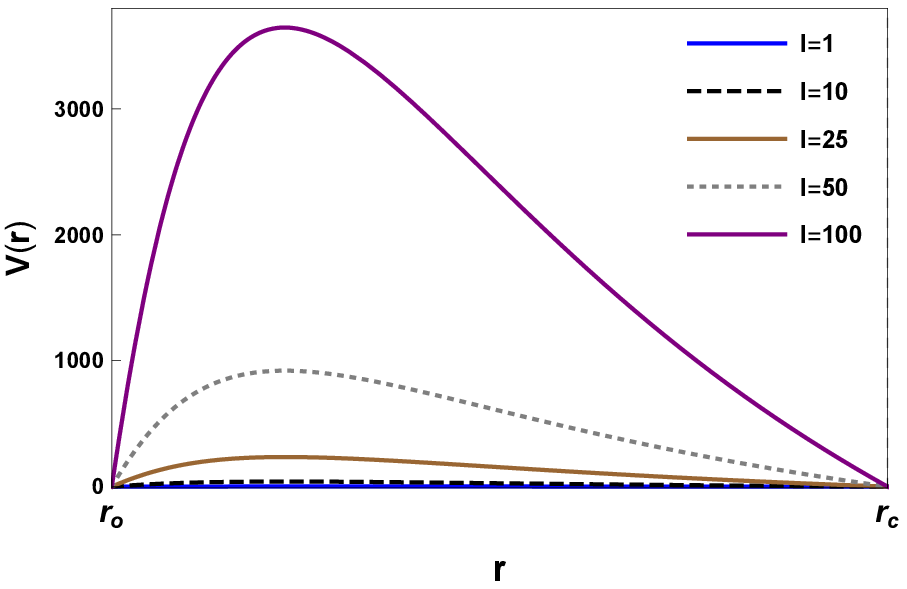}} \,
\centering
\subfigure[{Effective potentials in $l = 1$ and $\mu = 1$.}] {\includegraphics[scale=0.6,keepaspectratio]{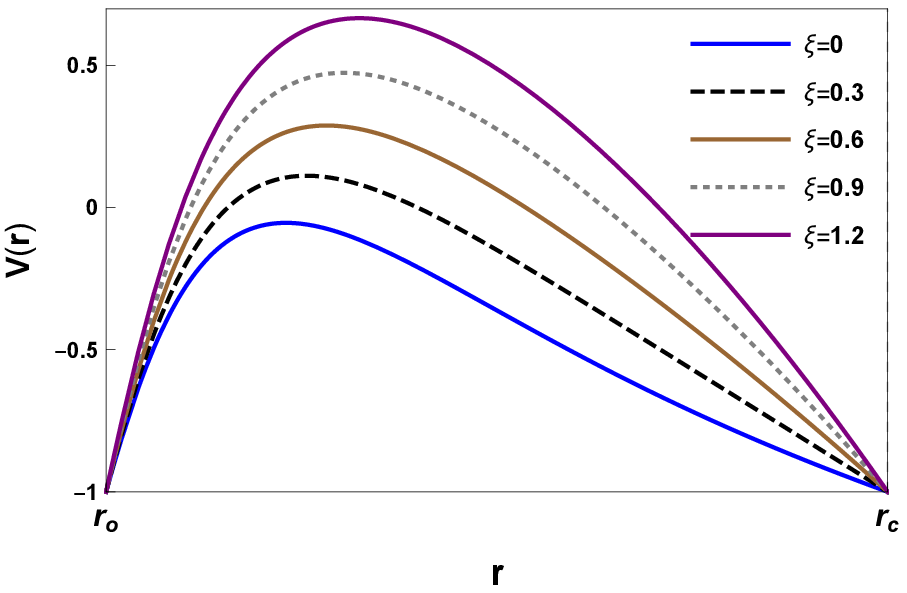}}\,
\centering
\subfigure[{Effective potentials in $l= 1$ and $\xi = 1$.}] {\includegraphics[scale=0.6,keepaspectratio]{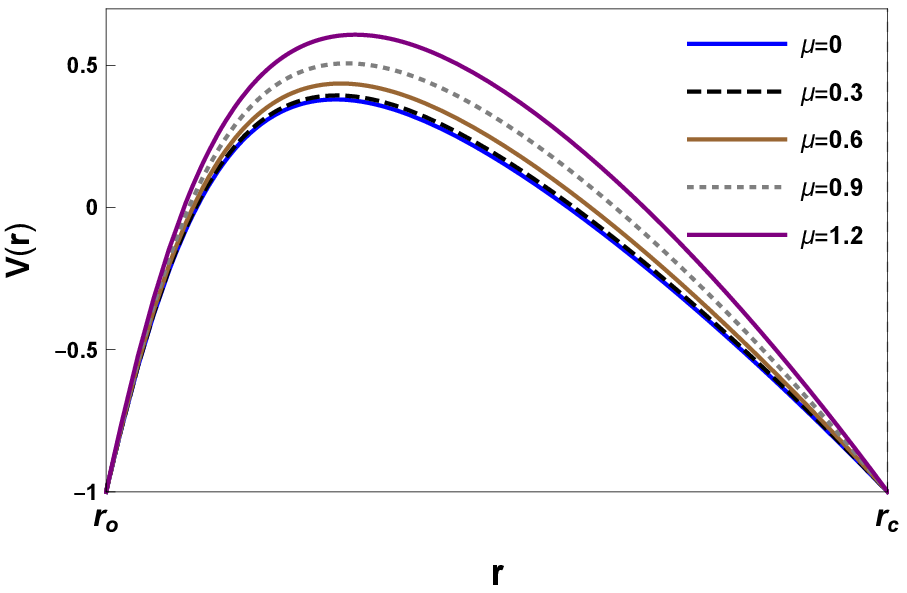}}
\centering
\caption{Effective potentials in $M=0.3$ and $\omega=1$ with respect to $l$, $\xi$, and $\mu$.}
\label{fig:quasi02}
\end{figure}

We consider the quasinormal resonance of non-minimally coupled massive scalar field in the MTZ black hole. In the SCC conjecture, the quasinormal frequency of the inward field plays an important role in estimating its decay rate at the outer horizon. Since the decay rate depends on the imaginary part of the frequency, we need to find a solution for the scalar field equation. Then, the action of the non-minimally coupled massive scalar field $\Psi$ is\cite{Crispino:2013pya}
\begin{align}
S_\Psi = -\frac{1}{2} \int d^4 x \sqrt{-g} (\partial_\mu \Psi \partial^\mu \Psi^* + (\mu^2+\xi \mathcal{R})\Psi \Psi^*),
\end{align}
where the mass of the scalar is $\mu$, and non-minimal coupling constant is $\xi$. Then, we can obtain the field equation with mass and non-minimal coupling terms.
\begin{align}\label{eq:scalareom01}
\frac{1}{\sqrt{-g}}\partial_\mu ( \sqrt{-g} g^{\mu\nu} \partial_\nu \Psi)-(\mu^2+\xi \mathcal{R})\Psi=0.
\end{align}
The solution to the scalar field $\Psi(t,r,\theta,\phi)$ is easily obtained in a simple form from Eq.(\ref{eq:scalareom01}). Then,
\begin{align}
\Psi(t,r,\theta,\phi)=\frac{e^{-i\omega t} e^{im\phi}}{r} R(r) Y_{l m}(\theta),
\end{align}
where $Y_{l m}$ is spherical harmonics. Further, $\omega$, $m$, and $l$ are separate variables corresponding to frequency and eigenvalues with respect to rotating axis and total angular momenta. Hence, the only non-trivial equation is the radial part, which is written as
\begin{align}\label{eq:radialeq}
\frac{1}{r^2}\partial_r(\Delta \partial_r R)+\left(\frac{r^2\omega^2}{\Delta}-\frac{l(l+1)}{r^2}-(\mu^2+\xi \mathcal{R})\right)R=0.
\end{align}
The radial equation in Eq.\,(\ref{eq:radialeq}) then becomes a Schr\"{o}dinger-like equation in a tortoise coordinate $r^*$, which is defined as
\begin{align}\label{eq:tortoise02}
\frac{d r^*}{d r}=\frac{r^2}{\Delta},
\end{align} 
where the range of the tortoise coordinate corresponds to
\begin{align}
r_\text{o}<r<r_\text{c} \rightarrow -\infty< r^* < \infty.
\end{align}
Under the tortoise coordinate, the radial equation is obtained as
\begin{align}\label{eq:potential}
\frac{d^2 R}{d {r^*}^2}+V R =0,\quad V=\omega^2 -\frac{\Delta}{r^4}\left( l(l+1)+r^2 (\mu^2+\xi \mathcal{R})-\frac{2M^2}{r^2}+\frac{2M}{r}-\frac{2\Lambda r^2}{3}\right).
\end{align}
The potential term determines detailed propagation of the scalar field. As shown in Fig.\,\ref{fig:quasi02}, there exists a peak between the outer and cosmological horizons.
Further, effective potentials in various parameters are similar to one another. At the spacetime boundary in the tortoise coordinate $r^*=\pm \infty$, the solutions to the radial equation in Eq.\,(\ref{eq:potential}) take on simpler forms because the boundaries correspond to outer and cosmological horizons satisfying $\Delta=0$.
\begin{align}
R(r)=e^{\pm i \omega r^*}\,\,\text{at} \,\, r^*\rightarrow \pm \infty.
\end{align}
Here, we impose boundary conditions for the quasinormal resonance, which is given by
\begin{align}\label{eq:boundarycondition01}
 &\Psi(t,r,\theta,\phi)=\frac{e^{-i\omega (t+r^*)} e^{im\phi}}{r} Y_{l m}(\theta) \,\, {\rm at} \,\, r^*\rightarrow -\infty,\\
 &\Psi(t,r,\theta,\phi)=\frac{e^{-i\omega (t-r^*)} e^{im\phi}}{r} Y_{l m}(\theta) \,\, {\rm at} \,\, r^*\rightarrow \infty,\nonumber
\end{align}
where a certain quasinormal mode is expected to originate from null geodesics in unstable circular orbits called the photon sphere\cite{Cardoso:2017soq}. In the SCC conjecture, whether or not the flux of the scalar field diverges is important, which is governed by the imaginary part of the frequency $\omega$ obtained in the near-horizon regime. This will be explored in the subsequent section.

\section{Strong Cosmic Censorship Conjecture}\label{sec4}

Here, we investigate whether the SCC conjecture is valid with respect to the non-minimally coupled massive scalar field in the MTZ black hole. By scalar field scattering, the perturbation can be blueshifted as it comes close to the Cauchy horizon. The blueshift is given by the amplification rate, which is related to the surface gravity of the Cauchy horizon $\kappa_\text{i}$\cite{Brady:1998au}. Note that the scalar field also undergoes exponential decay, which is given as $|\Psi-\Psi_0|\sim e^{-\alpha t}$ with the spectral gap $\alpha$. Then, destabilizing the Cauchy horizon depends on the competition between amplification and decay with respect to the perturbation, owing to the scalar field\cite{Cardoso:2017soq}. Further, the competition is governed by a very simple parameter, $\beta\equiv\alpha/\kappa_\text{i}$\cite{Cardoso:2017soq}. According to the SCC conjecture discussed in \cite{Christodoulou:2008nj}, the parameter $\beta$ determines whether the energy of the scalar field at the Cauchy horizon is divergent\cite{Dias:2018ufh}. When $\beta<\frac{1}{2}$, amplification becomes dominant, due to which the blueshifted inward mode is able to destabilize the Cauchy horizon. In this case, the SCC conjecture becomes valid. On the contrary, if $\beta>\frac{1}{2}$, then the quasinormal modes are damped. Further, the Cauchy horizon is still stable. In this case, the SCC conjecture is invalid. Therefore, the SCC conjecture can be elucidated from the value of $\beta$. In the following subsections, we consider the MTZ black hole in the limits of the large mass and small mass, and these limits are interpolated by the WKB approximation as given in \cite{Hod:2018dpx,Iyer:1986np,Iyer:1986nq}.

\subsection{Large Mass Case: Near-Extremal Black Holes}\label{sec41}

We investigate the imaginary part of the lowest frequency in the quasinormal resonance, which governs the decay of the influx in the near-extremal case. For a given cosmological constant, the maximum mass is achieved at the extremal case; thus, the massive limit implies the near-extremal case. In order to obtain the lowest frequency representing the dominant mode, the potential term in Eq.\,(\ref{eq:potential}) should be taken in the first order under the near-horizon limit of the potential peak. Then, the potential term needs to be written in terms of the tortoise coordinate from Eq.\,(\ref{eq:tortoise02})
\begin{align}\label{eq:tor1}
r^*=\frac{1}{2\kappa_\text{o}}\ln\left(\frac{r-r_{\rm o}}{r_{\rm c}-r}\right)+\frac{1}{2\kappa_\text{i}}\ln\left(\frac{r-r_{\rm n}}{r-r_{\rm i}}\right) .
\end{align}
When the peak of the potential barrier $r_p$ is located in the near-horizon regime, we can assume that $r_{\rm p}= r_{\rm o}+\epsilon$, where $\epsilon \ll 1$, because the outer and cosmological horizons are close in the near-extremal limit. Then, the location of the peak is approximately written in terms of Eq.\,(\ref{eq:tor1}) with the tortoise coordinate\cite{Fernando:2015kaa}
\begin{align}\label{eq:peaktortoise05}
r^*_{\rm p} = \frac{1}{2\kappa_{\rm o}}\ln\left(\frac{r_{\rm p}-r_{\rm o}}{r_{\rm c}-r_{\rm p}}\right) +\frac{1}{2\kappa_{\rm i}}\ln\left(\frac{r_{\rm o}-r_{\rm n}}{r_{\rm o}-r_{\rm i}}\right) +{\cal O}(\epsilon),
\end{align}
where the most dominant term is the first term in Eq.\,(\ref{eq:peaktortoise05}). Then, the location of the peak in the tortoise coordinate can be rewritten in the radial coordinate as
\begin{align}
r_{\rm p} \approx \frac{r_{\rm o}+r_{\rm c}e^{2\kappa_{\rm o} r^*_{\rm p}}}{1+e^{2\kappa_{\rm o} r^*_{\rm p}}} \label{eq:r_by_r_tor} .
\end{align}
As we have already assumed that the MTZ black hole was near-extremal, we can conclude that $r_{\rm c} = r_{\rm o}+\delta$ where $\delta \ll 1$. Note that $\epsilon$ should be smaller than $\delta$ because the peak is located in between outer and cosmological horizons. Under near-horizon and near-extremal conditions, the potential term in Eq.\,(\ref{eq:potential}) is obtained as
\begin{align}\label{eq:potential102}
V(r_{\rm p}) &=\left. \omega^2 -\frac{\Delta(r_{\rm p})}{r_{\rm p}^2}\left( \frac{l(l+1)}{r_{\rm p}^2}+ (\mu^2+\xi \mathcal{R})+\frac{r_{\rm p}^2}{2 r_{\rm p} \Delta(r_{\rm p})}
\left.\frac{d}{dr^2}\left(\frac{\Delta}{r^2}\right)^2\right|_{r=r_{\rm p}}\right)\right|_{r_{\rm p}=r_{\rm o}+\epsilon}  \\
 &= \omega^2 -\frac{\Lambda(r_{\rm c}-r_{\rm o})^2(r_{\rm o}-r_{\rm i})(r_{\rm o}-r_{\rm n})}{12r_{\rm o}^2 \cosh^2(\kappa_\text{o} r^*_{\rm p})}\left( \frac{l(l+1)}{r_{\rm o}^2}+ (\mu^2+\xi \mathcal{R}) \right),
\end{align}
where we take the leading order of $\epsilon$ and $\delta$, so neither of them appears in Eq.\,(\ref{eq:potential102}). Then, the radial equation in Eq.\,(\ref{eq:potential}) becomes
\begin{equation}\label{eq:PTpotential01}
\frac{d^2 R}{d {r^*}^2}+\left( \omega^2 -\frac{V_0}{\cosh^2(\kappa_{\rm o} r^*_{\rm p})}\right)R =0,\quad V_0=\frac{\Lambda(r_{\rm c}-r_{\rm o})^2(r_{\rm o}-r_{\rm i})(r_{\rm o}-r_{\rm n})}{12 r_{\rm o}^2 }\left( \frac{l(l+1)}{r_{\rm o}^2}+ (\mu^2+\xi \mathcal{R}) \right).
\end{equation}
This type of potential in Eq.\,(\ref{eq:PTpotential01}) is called the P\"{o}schl-Teller potential, whose solution is known. According to \cite{Ferrari:1984zz}, the quasinormal frequency $\omega$ is
\begin{equation}\label{eq:frequencies50001}
\omega = \sqrt{V_0-\frac{\kappa_\text{o}^2}{4}}- i \left(n+\frac{1}{2}\right) \kappa_\text{o}.
\end{equation}
The most dominant mode of the quasinormal resonance is at the least damping. This implies that the imaginary part of the dominant mode has the smallest value in all the possible values of $\text{Im}(\omega_n)$. With this, we can now determine the imaginary part of the dominant mode
\begin{align}\label{eq:ptdominantmodes01}
\text{Im}(\omega_{n=0})=\frac{1}{2}\kappa_\text{o}.
\end{align}
Therefore, the inequality of $\beta$ is obtained as
\begin{align}\label{eq:extremalscc01}
\frac{\text{Im}(\omega_{n=0})}{\kappa_\text{i}} < \frac{1}{2},
\end{align}
where we consider $\kappa_\text{o} < \kappa_\text{i}$. As a comparison of the decay and amplification rates of the quasinormal perturbation, the value of $\beta$ governs the stability of the Cauchy horizon, which plays a significant role in the SCC conjecture. The inequality in Eq.\,(\ref{eq:extremalscc01}) implies that the amplification is dominant in the near-extremal MTZ black hole of $r_\text{o}\approx r_\text{c}$. Then, the Cauchy horizon becomes unstable due to the blueshifted inward mode. Therefore, the SCC conjecture is valid for the massive MTZ black hole case.

\subsection{Small Mass Case: de Sitter Mode Approximation}\label{sec42}

As the mass of the MTZ black hole decreases, the size of the black bole also decreases. Finally, when the mass becomes zero, the geometry becomes the dS spacetime, containing only the cosmological horizon. Thus, we expect that the quasinormal modes physically smoothly become pure dS spacetime modes in the limit of the small mass. This behavior was already found in \cite{Cardoso:2017soq}, which dealt with the RNdS black hole in a massless scalar field, and was called a dS mode. We rewrite this behavior in terms of our notation and obtain the quasinormal mode of the non-minimally coupled massive scalar field in pure dS spacetime by modifying the result obtained in \cite{Du:2004jt}, which studied the massive scalar case.

As the mass tends to zero, the metric in Eq.\,(\ref{eq:metric01}) approximately becomes that of the dS spacetime. Hence, the potential term in the dS case is obtained from Eq.\,(\ref{eq:potential}) taken to the limit of $M$ and tending to zero. Then, the potential term becomes
\begin{equation}\label{eq:potential2}
V(r^*) = \omega^2+\frac{2\Lambda-3(\mu^2+\xi \mathcal{R})}{3 \cosh^2 \left(\sqrt{\frac{\Lambda}{3}}r^*\right)}-\frac{l(l+1)\Lambda}{3 \sinh^2\left(\sqrt{\frac{\Lambda}{3}}r^*\right)}.
\end{equation}
By substituting $\zeta=\cosh^{-2}\left(\sqrt{\frac{\Lambda}{3}}r^*\right)$ into Eq.\,(\ref{eq:potential2}), the radial equation is rewritten as
\begin{equation}\label{eq:DiffeqR}
\zeta(1-\zeta)\frac{d^2R}{d\zeta^2}+\left(1-\frac{3}{2}\zeta\right)\frac{dR}{d\zeta}+\frac{1}{4}\left(\frac{\omega^2\ell^2}{\zeta}-\frac{l(l+1)}{1-\zeta}-{\frac{3}{\Lambda}}(\mu^2+\xi \mathcal{R})+2\right)R=0.
\end{equation}
Note that $\zeta \rightarrow 1$ for $r^*\rightarrow 0$ ($r\rightarrow 0$) and $\zeta\rightarrow 0$ for $r^*\rightarrow \infty$ ($r\rightarrow r_c$). We take the ansatz to the radial function
\begin{align}
R(\zeta)=\zeta^k (1-\zeta)^p {\cal F}(\zeta).
\end{align}
Then, the radial equation in Eq.\,(\ref{eq:DiffeqR}) is rewritten as
\begin{align}\label{eq:hyper_eq}
&\zeta(1-\zeta)\frac{d^2\mathcal{F}}{d\zeta^2}+\left(1+2k-\left( 2k+2p+\frac{3}{2}\right)\zeta\right)\frac{d\mathcal{F}}{d\zeta}+\left[\frac{1}{\zeta}\left(k^2+\frac{\omega^2\ell^2}{4}\right)+\frac{1}{1-\zeta}\left(p^2-\frac{1}{2}p -\frac{1}{4}l(l+1)\right) \right. \nonumber\\
&\left. -\left((k+p)^2+\frac{1}{2}(k+p)+\frac{3}{4\Lambda}(\mu^2+\xi \mathcal{R})-\frac{1}{2}\right)\right]{\cal F}=0. 
\end{align}
Because the terms $1/\zeta$ and $1/(1-\zeta)$ in Eq.\,(\ref{eq:hyper_eq}) diverge at $\zeta=0$ or $\zeta=1$, they are eliminated by taking $k$ and $p$ as
\begin{equation}\label{eq:alphabeta}
k^2+\frac{3\omega^2}{4\Lambda}=0, \quad p^2-\frac{1}{2}p -\frac{1}{4}l(l+1)=0.
\end{equation}
Under the choice in Eq.\,(\ref{eq:alphabeta}), the radial equation in Eq.\,(\ref{eq:hyper_eq}) becomes the hypergeometric differential equation, whose general solution is given as
\begin{align}
R(\zeta) = A_0 \zeta^{-k}(1-\zeta)^p\,_2 {\cal F}_1 (a-c+1,b-c+1,2-c;\zeta) +  A_1 \zeta^{k}(1-\zeta)^p\,_2 {\cal F}_1 (a,b,c;\zeta),
\label{eq:Rsol1}
\end{align}
where
\begin{align}\label{eq:parametersabc1}
a &=k+p+\frac{1}{4}\left(1+\sqrt{1+4\left(2-\frac{3}{\Lambda}(\mu^2+\xi \mathcal{R})\right)}\right), \, b =k+p+\frac{1}{4}\left(1-\sqrt{1+4\left(2-\frac{3}{\Lambda}(\mu^2+\xi \mathcal{R})\right)}\right), \nonumber\\  c &=1+2k. 
\end{align}
Note that the general solution in Eq.\,(\ref{eq:Rsol1}) is assumed with respect to the non-integer $c$. This is a suitable form of the solution with arbitrary $\xi$ and $\mu$ of the scalar field. Instead of the non-integer $c$, the choice of the integer $c$ is still possible under a limited condition: $\mu^2+\xi \mathcal{R}=0$. The case of the integer $c$ is coincident with the massless scalar field without the coupling. This is already discussed in \cite{Cardoso:2017soq}, and our analysis of $\omega$ technically includes the massless case\cite{Du:2004jt}. Hence, to keep the arbitrary $\xi$ and $\mu$ for general cases, we focus on the solution about the non-integer $c$. According to the boundary condition for the quasinormal resonance in Eq.\,(\ref{eq:boundarycondition01}), the scalar field only has a purely outgoing mode near the cosmological horizon $\zeta \rightarrow 0$. In addition, the scalar field is assumed to have vanished at the origin of the spacetime, $\zeta =1$. Thus, we should take $A_0=0$ in order to eliminate the incoming wave at the cosmological horizon. This fixes $k=-i\ell\omega/2$. Then, the solution in Eq.~\eqref{eq:Rsol1} is reduced to
\begin{align}\label{eq:Rsol2}
R(\zeta) =A_1 \zeta^{- i\sqrt{\frac{3}{\Lambda}}\frac{\omega }{2}}(1-\zeta)^p\,_2 {\cal F}_1 (a,b,c;\zeta).
\end{align}
To impose the boundary condition at the origin, we can rewrite Eq.\,(\ref{eq:Rsol2}) under the transformation $\zeta\rightarrow 1-\zeta$ as
\begin{align}\label{eq:Rsol3} 
R(\zeta)=&A_1 \left[\zeta^{- i \sqrt{\frac{3}{\Lambda}}\frac{\omega }{2}} (1-\zeta)^p \frac{\Gamma(c)\Gamma(c-a-b)}{\Gamma(c-a)\Gamma(c-b)}\, _2{\cal F}_1(a,b,a+b-c+1;1-\zeta) \right. \notag \\
&\left. +\zeta^{- i \sqrt{\frac{3}{\Lambda}}\frac{\omega }{2}} (1-\zeta)^{\frac{1}{2}-p}\frac{\Gamma(c)\Gamma(a+b-c)}{\Gamma(a)\Gamma(b)}\, _2{\cal F}_1(c-a,c-b,c-a-b;1-\zeta) \right]. 
\end{align}
According to Eq.\,(\ref{eq:alphabeta}), there are two solutions for $p$ in Eq.\,(\ref{eq:alphabeta}): $p = -l/2$ or $p = (1+l)/2$. (a) When we choose $p = -l/2$, the parameters in Eq.\,(\ref{eq:parametersabc1}) are fixed as $c-a=-n$ or $c-b=-n$, because the radial solution should be regular at the origin $\zeta\rightarrow 1$. Then, the radial solution in Eq.\,(\ref{eq:Rsol2}) becomes
\begin{align}
R(\zeta)=&A_1 \left[\zeta^{- i \sqrt{\frac{3}{\Lambda}}\frac{\omega }{2}} (1-\zeta)^{\frac{1}{2}(1+l)}\frac{\Gamma(c)\Gamma(a+b-c)}{\Gamma(a)\Gamma(b)}\, _2{\cal F}_1(c-a,c-b,c-a-b;1-\zeta) \right].
\end{align}
(b) When we choose $p=(1+l)/2$, the parameters in Eq.\,(\ref{eq:parametersabc1}) are fixed as  $a=-n$ or $b=-n$ to be a regular radial solution. Then, the radial solution in this choice is 
\begin{align}
R(\zeta)=&A_1 \left[\zeta^{- i \sqrt{\frac{3}{\Lambda}}\frac{\omega }{2}} (1-\zeta)^{\frac{1}{2}(1+l)} \frac{\Gamma(c)\Gamma(c-a-b)}{\Gamma(c-a)\Gamma(c-b)}\, _2{\cal F}_1(a,b,a+b-c+1;1-\zeta)  \right].
\end{align}
The main concern of this study is the frequency $\omega$ rather than the radial solutions. Interestingly, in combination with Eq.\,(\ref{eq:parametersabc1}), the frequencies in both choices of $p = -l/2$ and $p = (1+l)/2$ are exactly coincident to
\begin{equation}\label{eq:dsmodes201}
\omega = -i\sqrt{\frac{\Lambda}{3}}\left(2n+l+\frac{3}{2}\pm \sqrt{\frac{9}{4}+12 \xi-\frac{3}{\Lambda}\mu^2 }\right),
\end{equation}
which is given as a pure imaginary, which means all the modes will decay. The most dominant mode among them is the one that is least damping in Eq.\,(\ref{eq:dsmodes201}). Thus, the dominant mode is in $n=0$ and $l=1$ with the choice of the minus sign. Then, the quasinormal frequency is in the dS spacetime
\begin{equation}\label{eq:omega_dS}
\omega^{\text{dS}}_{n=0} = -i\sqrt{\frac{\Lambda}{3}}\left(\frac{5}{2}- \sqrt{\frac{9}{4}+12 \xi-\frac{3}{\Lambda}\mu^2 }\right).
\end{equation}
Note that the most dominant mode of the minimally-coupled massless scalar field is given as $\omega_{n=0}=-il\kappa_\text{c}$ with $\mu=0$, $\xi=0$, $n=0$, and $l=1$. This is consistent with the dS mode in \cite{Cardoso:2017soq}. Here, we expect that the quasinormal mode in the MTZ black hole will come close to the pure dS mode of Eq.\,(\ref{eq:omega_dS}) in the zero-mass limit. Then, the rate between amplification and decay are obtained as
\begin{align}\label{eq:omegadSbeta1}
\lim_{M\rightarrow 0}\beta=\frac{|\text{Im}(\omega^{\text{dS}}_{n=0})|}{{\lim}_{M\rightarrow 0}\kappa_\text{i}}= \left(\frac{5}{2}- \sqrt{\frac{9}{4}+12 \xi-\frac{3}{\Lambda}\mu^2 }\right),
\end{align}
where we impose $\lim_{M\rightarrow 0}\kappa_\text{i}=\sqrt{\frac{\Lambda}{3}}$ in the zero-mass limit and under the lukewarm condition. Therefore, the value of $\beta$ only depends on the mass and non-minimal coupling constant of the scalar field in Eq.\,(\ref{eq:omegadSbeta1}). In the choice of $\mu=0$ and $\xi=0$, it is easily shown that $\beta=1$, which implies the violation of the SCC conjecture, is shown in \cite{Cardoso:2017soq}.
\begin{figure}[h]
\begin{center}
\includegraphics[scale=0.75,keepaspectratio]{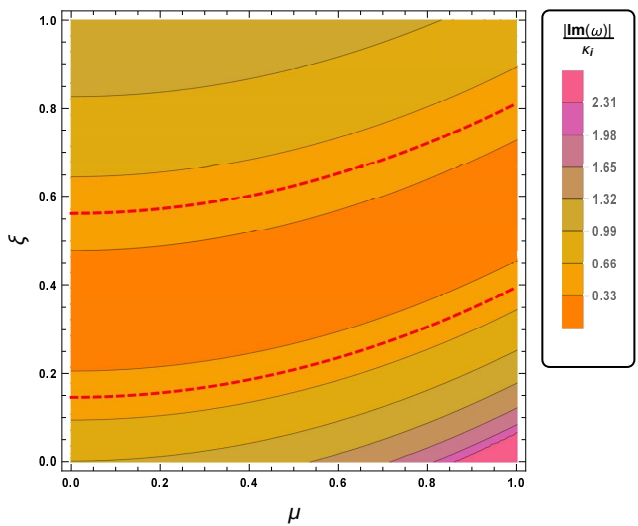}
\end{center}
\caption{$\beta$ of the dominant mode of dS spacetime with $\Lambda=1$.}\label{fig:quasi_dS} 
\end{figure}
However, under the quasinormal resonance of the non-minimally coupled massive scalar field, other choices can be possible, such as
\begin{align}\label{eq:validsccsmallmass01}
12\xi-\frac{3}{\Lambda}\mu^2>\frac{7}{4}.
\end{align}  
Then, $\beta$ can be smaller than $\frac{1}{2}$. Therefore, the SCC conjecture can be valid in the range shown in Eq.\,(\ref{eq:validsccsmallmass01}). More detailed behaviors are shown in Fig.\,\ref{fig:quasi_dS}. In the small mass and coupling constant of the scalar field, $\beta$ exceeds $\frac{1}{2}$, so the amplification can be dominant instead of decay rate. However, there are ranges of $\mu$ and $\xi$ that give a $\beta$ smaller than $\frac{1}{2}$, between the red dashed lines in Fig.\,\ref{fig:quasi_dS}. Here, the decay of the scalar field is more efficient than the amplification. Thus, the Cauchy horizon can be stable during the perturbation. Note that this study only considers the linear perturbation of the quasinormal resonance; thus, a detailed discussion is needed to consider other effects such as self-interaction and backreaction of the scalar field.

\subsection{WKB Approximation}\label{sec43}

\begin{figure}[h]
\centering
\subfigure[{$\text{Re}(\omega)$ for $l=1$.}] {\includegraphics[scale=0.9,keepaspectratio]{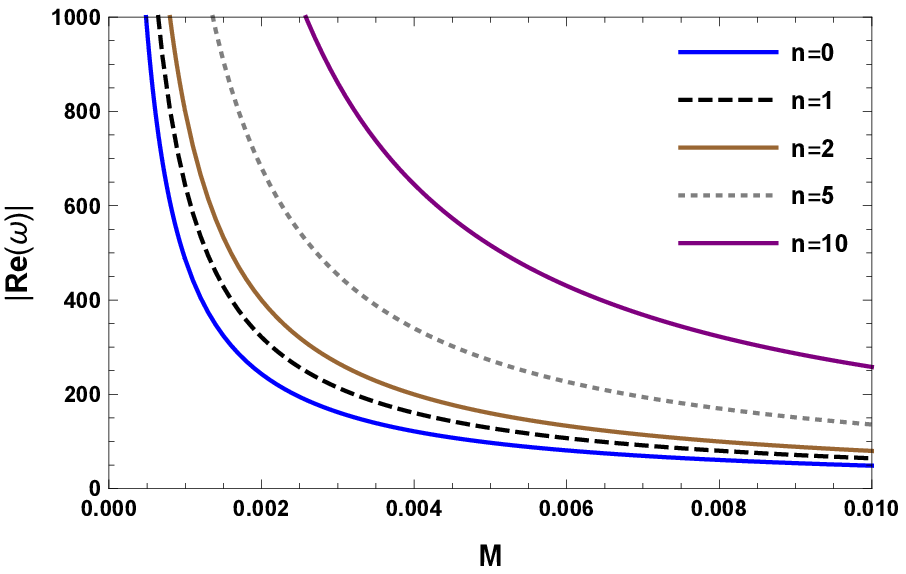}} \quad
\subfigure[{$\text{Im}(\omega)$ for $l=1$.}] {\includegraphics[scale=0.9,keepaspectratio]{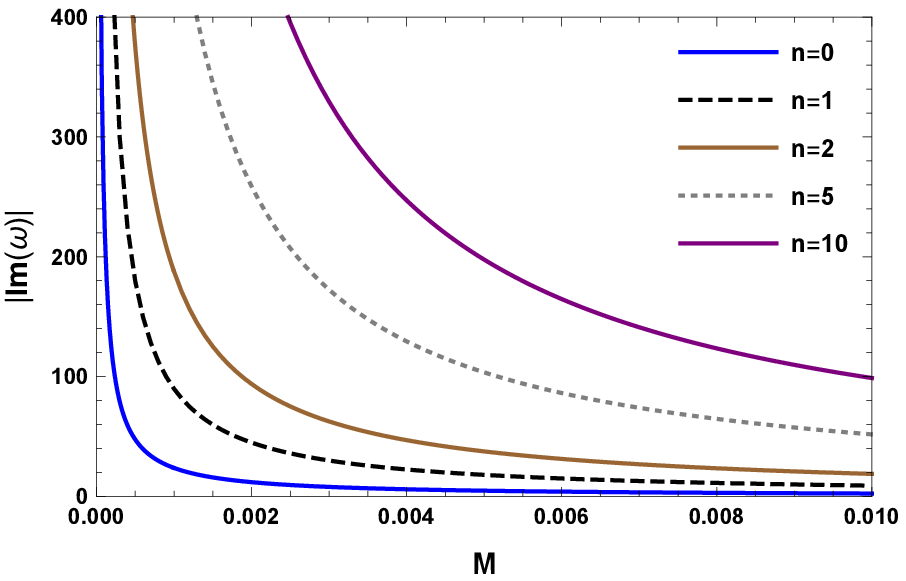}}
\centering
\subfigure[{$\text{Re}(\omega)$ for $n=0$.}] {\includegraphics[scale=0.9,keepaspectratio]{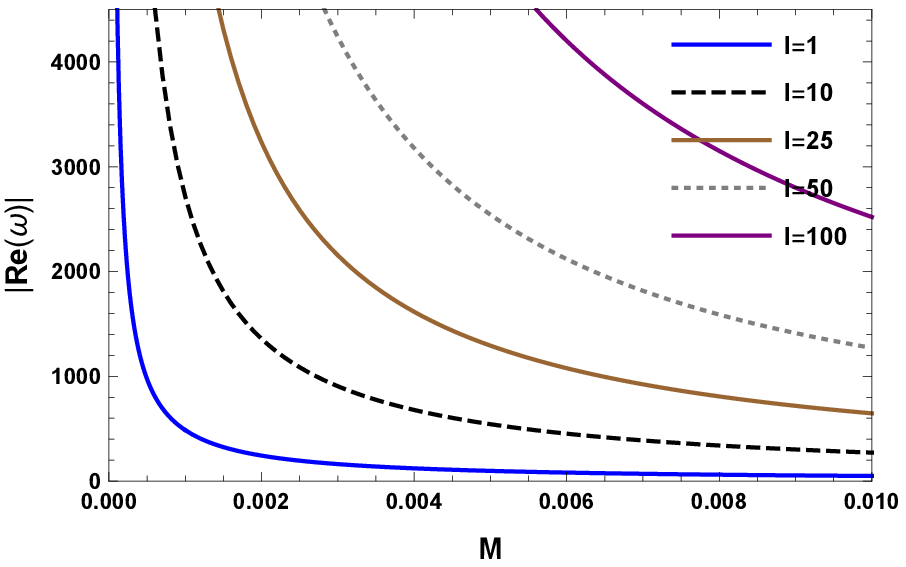}} \quad
\subfigure[{$\text{Im}(\omega)$ for $n=0$.}] {\includegraphics[scale=0.88,keepaspectratio]{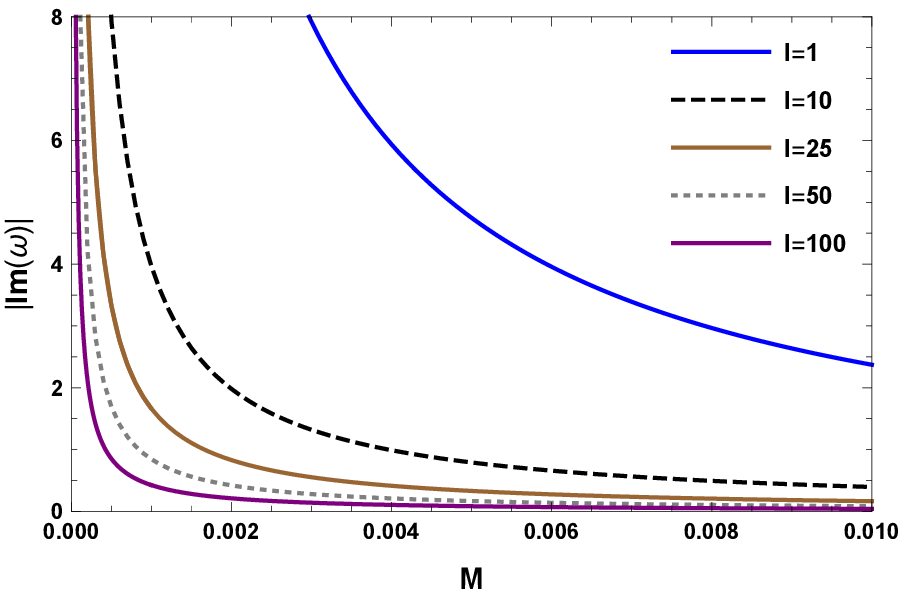}}
\caption{{\small Quasinormal modes with respect to the mass of the MTZ black hole, for $\xi=1,~\mu=1$.}}
\label{fig:reandimfreq05}
\end{figure}

The SCC conjecture is discussed in the large- and small-mass limits of the MTZ black hole. However, the validity of the SCC conjecture still remains in between the two limits, according to our analysis. To interpolate these limits, we apply the WKB approximation to obtain $\omega_{n=0}$ in the eikonal limit $l\gg 1$ with respect to the mass of the MTZ black hole. The WKB approximation, which will be used here, is given in \cite{Iyer:1986np,Iyer:1986nq}. Then, the quasinormal frequencies of Eq.\,(\ref{eq:potential}) are determined to\cite{Iyer:1986nq}
\begin{align}\label{eq:potentialvstar1}
V_\text{p}^*=\sqrt{-2V_\text{p}^{*(2)}}\Lambda_{n=0}-\frac{i}{2}\sqrt{2V_\text{p}^{*(2)}}\left(1+\Omega_{n=0}\right),\quad V^*_\text{p}\equiv V(r^*)\Big|_{r^*=r^*_\text{p}},\quad V^{*(n)}_\text{p}\equiv \frac{d^{n}V(r^*)}{(dr^{*})^n}\Big|_{r^*=r^*_\text{p}},
\end{align} 
where
\begin{align}
\Lambda_{n=0}=\frac{1}{\sqrt{2V_\text{p}^{*(2)}}}&\left(\frac{1}{16}\left(\frac{V^{*(4)}_\text{p}}{V^{*(2)}_\text{p}}\right)-\frac{11}{144}\left(\frac{V^{*(3)}_\text{p}}{V^{*(2)}_\text{p}}\right)^2\right), 
\end{align}
\begin{align}
\Omega_{n=0}=\frac{1}{2V_\text{p}^{*(2)}}&\left(\frac{155}{1728}\left(\frac{V^{*(3)}_\text{p}}{V^{*(2)}_\text{p}}\right)^4\right.-\frac{19}{96}\left(\frac{\left({V^{*(3)}_\text{p}}\right)^2V^{*(4)}_\text{p}}{\left({V^{*(2)}_\text{p}}\right)^3}\right)+\frac{7}{192}\left(\frac{V^{*(4)}_\text{p}}{V^{*(2)}_\text{p}}\right)^2 \notag \\
&\left.+\frac{13}{144}\left(\frac{{V^{*(3)}_\text{p}}V^{*(5)}_\text{p}}{\left({V^{*(2)}_\text{p}}\right)^2}\right)-\frac{1}{48}\left(\frac{V^{*(6)}_\text{p}}{V^{*(2)}_\text{p}}\right)\right).\nonumber
\end{align}
$V_\text{p}^*$ only includes the quasinormal frequency $\omega$ in Eq.\,(\ref{eq:potentialvstar1}), we can obtain the quasinormal frequencies for the masses of the MTZ black hole as shown in Fig.\,\ref{fig:reandimfreq05}. 
Each mode is represented by a pair of $\text{Re}(\omega)$ and $\text{Im}(\omega)$ with the same color. The real part of the quasinormal frequency is about a propagating oscillation in Fig.\,\ref{fig:reandimfreq05}\,(a) and (c). The decay rate is closely related to the imaginary part of $\text{Im}(\omega)$ in Fig.\,\ref{fig:reandimfreq05}\,(b) and (d). Here, we need to determine the dominant mode, which is the least damping mode, so that it will have the longest life span among all the modes. In Fig.\,\ref{fig:reandimfreq05}\,(b), the mode with $n=0$ is in the least damping state for various values of $n$. Further, for a fixed $n$, the least damping mode appears at the largest value of $l$ in Fig.\,\ref{fig:reandimfreq05}\,(d). Particularly, the damping becomes smaller as $l$ increases. Thus, for our analysis, we assume the eikonal limit, ($l\gg 1$) to $l=100$. Therefore, the dominant mode in the quasinormal resonance is in $n=0$ and $l=100$. Note that the effects of $\mu$ and $\xi$ do not much affect our analysis; therefore, we did not introduce them in Fig.\,\ref{fig:reandimfreq05}.

\begin{figure}[h]
\centering
\subfigure[{Values of $\beta$ in $\xi=1$ and $\mu=1$.}] {\includegraphics[scale=0.9,keepaspectratio]{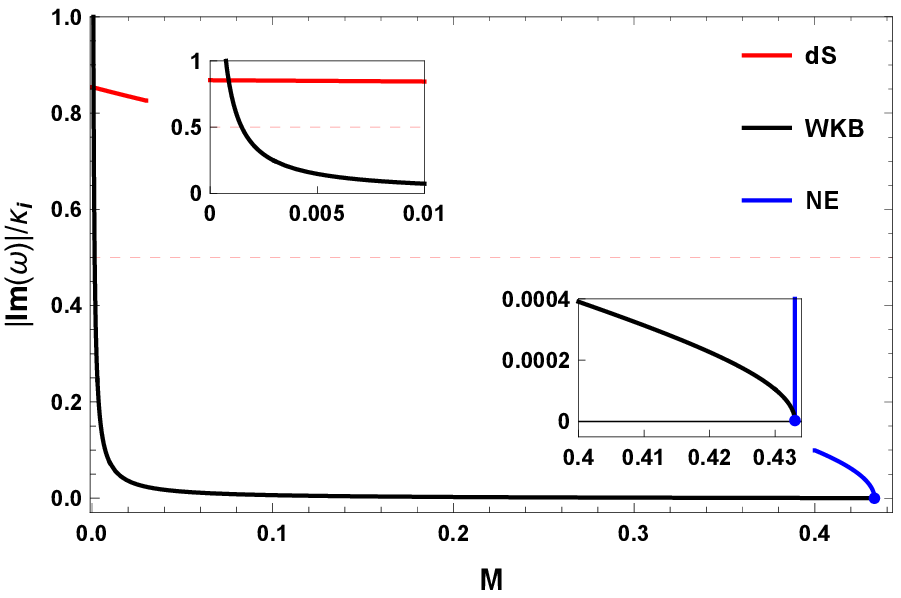}} \quad
\centering
\subfigure[{Values of $\beta$ in $\xi=0.5$ and $\mu=0.5$.}] {\includegraphics[scale=0.9,keepaspectratio]{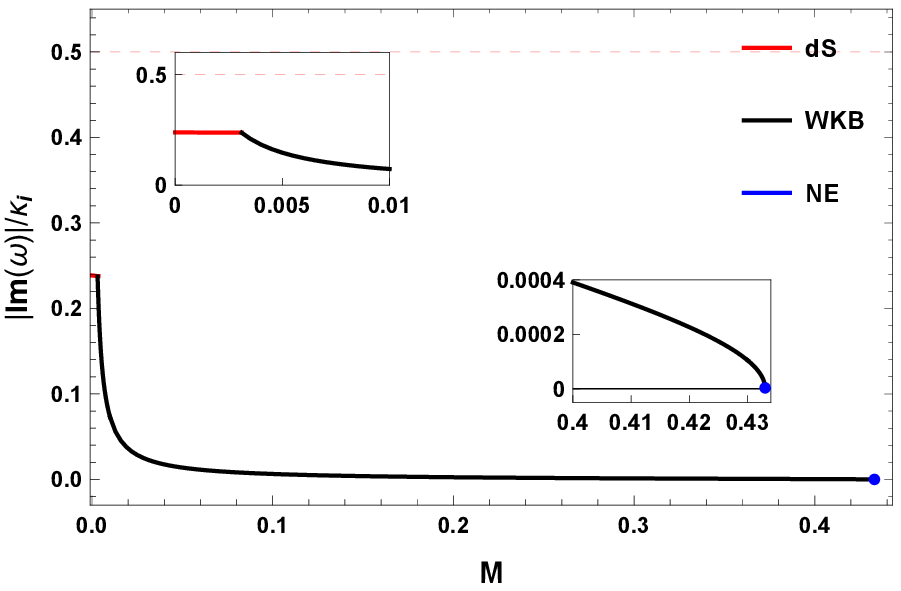}}
\caption{{\small Values of $\beta$ of quasinormal modes with two limits of MTZ black holes.}}
\label{fig:quasi0100a}
\end{figure}

We can now integrate all our results for the $\beta$ of quasinormal modes in the MTZ black holes, and show them in Fig.\,\ref{fig:quasi0100a} with magnified graphs in two limits.
The detailed behaviors of $\beta$ with two limits are given in Fig.\,\ref{fig:quasi0100a}\,(a). The dominant modes obtained from the WKB approximation is represented by the black lines. There are two limits represented by blue and red lines. The blue line is for the $\beta$ obtained from the P\"{o}schl-Teller potential in the large-mass limit of the near-extremal approximation in Eq.\,(\ref{eq:ptdominantmodes01}). At this limit, we can clearly observe that the WKB and P\"{o}schl-Teller potential approximations are exactly coinciding as represented by the blue point. Further, the value of $\beta$ is much smaller than $\frac{1}{2}$. Thus, the SCC conjecture is valid for this limit. However, in the small-mass limit of the MTZ black hole, the behavior patterns become complex. The value of $\beta$ in the quasinormal resonance rises to the infinity in the limit of the small mass in the black hole. On the other hand, the value of $\beta$ of the dominant quasinormal resonance in the dS spacetime is finite and much smaller than that of the black hole. As we already expected for the small-mass limit in Sec.\,\ref{sec42}, the dominant mode of the quasinormal resonance of the black hole, denoted by a black line, may smoothly converge on that of the dS spacetime, denoted by a red line in Fig.\,\ref{fig:quasi0100a}\,(a). This is because, in the limit of the small mass, the black hole could be too small, the configuration of the quasinormal resonance may be a superposition of those resonances of black hole and dS spacetime. Then, the quasinormal resonance of the dS spacetime can be dominant as it has a smaller $\beta$ value than the black hole. Different from the WKB approximation, in the small-mass limit, $\beta$ has a finite value depending on $\xi$ and $\mu$ in Eq.\,(\ref{eq:omegadSbeta1}). In Fig.\,\ref{fig:quasi0100a}\,(a), there is a crossing point of red and black lines where the dominant mode transforms into that of the dS spacetime $l=1$ rather than into the eikonal limit $l=100$. Hence, the dominant quasinormal resonance is assumed to be the red line of the dS spacetime in the mass smaller than the crossing point. The small-mass limit shows that $\beta$ can rise over $\frac{1}{2}$ at the small mass. Therefore, the SCC conjecture may be invalid in the small-mass range. However, there exists a specific case where the SCC conjecture is valid for all the masses of the MTZ black hole, as shown in Fig.\,\ref{fig:quasi0100a}\,(b). Here, we simplify the diagram by choosing the smaller $\beta$ for a given mass, because the dominant mode should have the smallest damping factor, $\text{Im}(\omega)$. The blue point is still largely coincident with the WKB approximation. Interestingly, according to the choice of $\xi$ and $\mu$, $\beta$ can be lower than $\frac{1}{2}$ at the small-mass limit as shown in Fig.\,\ref{fig:quasi0100a}\,(b). This implies that the SCC conjecture is valid for all the masses of the MTZ black hole.

We have applied various methods to obtain the dominant modes of the quasinormal resonance, including approximate potentials and WKB method, in MTZ black holes. Our results are consistent with previous studies conducted on the massless scalar field. We will now check the consistency of the results with those of the previous studies that discuss RNdS black holes. As we consider the quasinormal mode of the scalar field in the MTZ black hole, our analysis needs to be consistent with \cite{Cardoso:2017soq}. Note that the detailed method is different. The MTZ black hole can be considered for the lukewarm case of $Q/M=1$ compared with the RNdS black hole in \cite{Cardoso:2017soq}. For the near-extremal black hole of $r_\text{o}\approx r_\text{c}$, the SCC conjecture is valid because $\beta < \frac{1}{2}$. This is consistent with our results given in Sec.\,\ref{sec41}. Moreover, for the small-mass limit of $r_\text{o},\,r_\text{i}\rightarrow 0$, the value of $\beta$ becomes larger than $\frac{1}{2}$, which means that the conjecture is invalid. This is also consistent with our results in Sec.\,\ref{sec42}, in the case of the massless scalar field of $\xi=0$ and $\mu=0$. Then, we can find a point  $\beta=\frac{1}{2}$ after interpolating the two limits by the WKB approximation. This is also provided in Sec.\,\ref{sec43}. Although MTZ and lukewarm RNdS black holes are solutions to different theories of gravity, their quasinormal frequencies are consistent with one another. Whether or not non-linear effect is subtle is still an issue in current studies. Our analysis is based on the quasinormal resonance of a linear perturbation, and does not consider non-linear effects. The non-linear effect is a subject for future study.

\section{Summary}\label{sec6}

We investigated the validity of the SCC conjecture in the MTZ or lukewarm RNdS black hole by the quasinormal resonance of the non-minimally coupled massive scalar field. Since the instability of the Cauchy horizon depends on the amplification and decay rates of the quasinormal resonance, we obtained the overall behaviors of $\beta\equiv \text{Im}(\omega)/\kappa_\text{i}$ with respect to the mass of the black holes. In the analysis of the SCC conjecture by quasinormal modes, the dominant mode, which is the least damping mode, plays an important role. Therefore, we first obtained the values of $\beta$ for the small- and large-mass limits of the MTZ black hole. Then, we combined them by the WKB approximation. In the large-mass limit, for a given cosmological constant, the black hole becomes the near-extremal case where the effective potential of the scalar field reduces to the P\"{o}schl-Teller potential. Then, we obtained the quasinormal frequency and read the value of $\beta<\frac{1}{2}$ for the dominant mode. This implies that the amplification of the inward field is more dominant than the decay rate. Hence, the SCC conjecture is valid for large mass black holes. Note that as the mass of the black hole decreases, the value of $\beta$ tends to increase. Particularly, at the small-mass limit, $\beta$ rapidly diverges in the eikonal limit of the scalar field. To resolve this divergence, we carefully consider that the dominant mode of the quasinormal resonance gradually becomes similar to that of the dS spacetime, because the black hole could be too small to affect the quasinormal mode in the limit. Hence, it is possible that the least damping mode originates from that of the dS spacetime. Moreover, the dominant mode of the dS spacetime had less damping than that found by the WKB approximation. Thus, the amplification and decay rates depend on the dS spacetime mode in the small-mass limit. Instead of the mass of the black hole, the dominant mode of the dS spacetime is determined precisely from $\xi$ and $\mu$. In the limit of the massless scalar field without coupling, as shown in previous studies, the value of $\beta$ is still larger than $\frac{1}{2}$. On the contrary, we found that there is a range producing smaller $\beta$ than $\frac{1}{2}$ in the phase-space of $\xi$ and $\mu$. This implies that the SCC conjecture is invalid. Therefore, the validity of the SCC conjecture depends on the coupling constant and mass of the scalar field in the lukewarm RNdS or MTZ black hole. It should be noted that our analysis is based on the perturbations of the quasinormal mode. Therefore, it can be improved by considering backreaction or non-linear effects. Nevertheless, the results of this study extend the investigation of the SCC conjecture to the non-minimally coupled massive scalar field. Further, we found that the validity of the SCC conjecture in this case may be different from that of the massless scalar field without coupling.

\vspace{10pt} 

\noindent{\bf Acknowledgments}

\noindent This work was supported by the National Research Foundation of Korea (NRF) grant funded by the Korea government (MSIT) (NRF-2018R1C1B6004349). B.G. would like to thank Yongwan Gim for helpful discussions. In addition, B.G. appreciates APCTP for its hospitality during completion of this work.

\end{document}